\documentclass[conference]{IEEEtran}
\IEEEoverridecommandlockouts

\usepackage{cite}
\usepackage{amsmath,amssymb,amsfonts}
\usepackage{algorithmic}
\usepackage{graphicx}
\usepackage{textcomp}
\usepackage{xcolor}

\usepackage{amsmath, calc}
\usepackage{graphicx}
\usepackage{tcolorbox}
\usepackage{multirow}
\usepackage{url}
\usepackage{hyperref}
\usepackage{pgfplots}
\pgfplotsset{compat=1.18}
\usepackage{textcomp}
\usepackage{placeins}
\usepackage{fontawesome}
\usepackage{comment}
\usepackage{tabularx}
\usepackage{enumitem}
\usepackage{booktabs}
\usepackage{comment}

\usepackage{hyperref}

\usepackage{framed}

\usepackage{tcolorbox}
\usepackage{fancybox} 

\usepackage{tabularx}

\usepackage{makecell}
\usepackage{multirow}
\usepackage{csquotes}
\usepackage{graphicx}
\usepackage{orcidlink}
\usepackage{multicol}
\usepackage{colortbl}
\usepackage{fancybox,framed}
\usepackage{soul}
\usepackage{subfig}

\def\BibTeX{{\rm B\kern-.05em{\sc i\kern-.025em b}\kern-.08em
    T\kern-.1667em\lower.7ex\hbox{E}\kern-.125emX}}
\begin{document}

\makeatletter
\newcommand{\linebreakand}{%
  \end{@IEEEauthorhalign}
  \hfill\mbox{}\par
  \mbox{}\hfill\begin{@IEEEauthorhalign}
}
\makeatother

\title{How to Elicit Explainability Requirements? A Comparison of Interviews, Focus Groups, and Surveys}


\author{
    \IEEEauthorblockN{
        Martin Obaidi\orcidlink{0000-0001-9217-3934},
        Jakob Droste\orcidlink{0000-0001-8746-6329},
        Hannah Deters\orcidlink{0000-0001-9077-7486}, \\
        Marc Herrmann\orcidlink{0000-0002-3951-3300},
        Raymond Ochsner,
        Kurt Schneider\orcidlink{0000-0002-7456-8323}
    }
    \IEEEauthorblockA{
        \textit{Leibniz Universität Hannover, Software Engineering Group} \\
        Hannover, Germany \\
        \{martin.obaidi, jakob.droste, hannah.deters\}@inf.uni-hannover.de, \\
        \{marc.herrmann, kurt.schneider\}@inf.uni-hannover.de, \\ raymond.ochsner@stud.uni-hannover.de
    }
    \and
    \IEEEauthorblockN{
        Jil Klünder\orcidlink{0000-0001-7674-2930}
    }
    \IEEEauthorblockA{
        \textit{University of Applied Sciences} \\
        \textit{FHDW Hannover} \\
        Hannover, Germany \\
        jil.kluender@fhdw.de
    }
}


\maketitle

\begin{abstract}
As software systems grow increasingly complex, explainability has become a crucial non-functional requirement for transparency, user trust, and regulatory compliance. Eliciting explainability requirements is challenging, as different methods capture varying levels of detail and structure. This study examines the efficiency and effectiveness of three commonly used elicitation methods—focus groups, interviews, and online surveys—while also assessing the role of taxonomy usage in structuring and improving the elicitation process. We conducted a case study at a large German IT consulting company, utilizing a web-based personnel management software. A total of two focus groups, 18 interviews, and an online survey with 188 participants were analyzed. The results show that interviews were the most efficient, capturing the highest number of distinct needs per participant per time spent. Surveys collected the most explanation needs overall but had high redundancy. Delayed taxonomy introduction resulted in a greater number and diversity of needs, suggesting that a two-phase approach is beneficial. Based on our findings, we recommend a hybrid approach combining surveys and interviews to balance efficiency and coverage. Future research should explore how automation can support elicitation and how taxonomies can be better integrated into different methods.
\end{abstract}


\section{Introduction}
\label{sec:intro}
As the complexity of modern software systems continues to increase, explainability is gaining importance as a non-functional requirement (NFR) in software engineering~\cite{kohl2019explainability,chazette2020explainability}. Explainability requirements may concern and affect different system aspects~\cite{chazette2021exploring,droste2024explanations}, such as privacy~\cite{brunotte2023context,brunotte2023privacy}, security~\cite{vigano2020explainable}, computer-human interaction~\cite{deters2024UXandExplainability,obaidi2025automatingexplanationneedmanagement}, or artificial intelligence~\cite{adadi2018peeking}. As such, an effective implementation of system-integrated explanations depends on the employment of appropriate requirements engineering techniques for implementing appropriate explanations~\cite{chazette2022can}. Those explainability requirements may be elicited via a variety of different methods~\cite{chazette2022can}. One approach is to analyze direct user feedback~\cite{anders2023userfeedback,anders2022userfeedback}, such as app store reviews, support channels like emails, or help-desk queries~\cite{obaidi2025automatingexplanationneedmanagement,unterbusch2023explanation}. Alternatively, more traditional methods such as surveys, interviews, and workshops or focus groups can be employed~\cite{chazette2022can}. Depending on the method, the elicited requirements differ in number, depth, and quality~\cite{zowghi2005requirements}. Existing research in security requirements engineering has revealed that different elicitation methods have distinct strengths~\cite{fabian2010comparison}. Related work in the context of explainability has shown that while methods like surveys and questionnaires yield quantitative data that provide an insightful overview, qualitative data from interviews or workshops provide the detailed feedback needed to implement explanations sensibly~\cite{droste2023designing}.

To facilitate the efficient handling of explainability requirements, developers can use taxonomies to classify user needs based on system-specific or general elements~\cite{droste2024explanations,unterbusch2023explanation}. Taxonomies can be useful tools in requirements elicitation, as they may help stakeholders understand what kinds of requirements exist and what they entail~\cite{droste2024explanations}. Furthermore, taxonomies may be used for automation in requirements management, as demonstrated in recent research~\cite{obaidi2025automatingexplanationneedmanagement}.

To investigate the efficiency and effectiveness of different elicitation methods for collecting explainability requirements, this paper examines three commonly used manual approaches: focus groups, interviews, and online surveys. In conducting this study, we utilize the explanation need taxonomy developed by Droste et al.~\cite{droste2024explanations} to structure and support the elicitation process. To understand how these methods compare to each other, we conducted a case study at a large German IT consulting company, utilizing a web-based personnel management software as the case. We used the three elicitation methods throughout multiple iterations with stakeholders at the case company and compared their effectiveness and efficiency in capturing explainability requirements. To gain additional insights into how taxonomy usage affects the completeness, structure, and diversity of elicited explanation needs, we investigated the impact of introducing the explainability need taxonomy during the data collection process.

In summary, the contributions of this work cover the following:
\begin{itemize}
    \item Empirical comparison of three elicitation methods --- focus groups, interviews, and surveys --- regarding their efficiency and effectiveness in capturing explainability requirements.
    \item Investigation of the impact of taxonomy usage on the elicitation process, analyzing whether and when categorizing explanation needs by type improves their identification.
    \item Evaluation of how elicitation methods influence the distribution of explanation needs across taxonomy categories, highlighting differences in the types of needs identified.
\end{itemize}

The rest of this paper is structured as follows: Section~\ref{sec:background} reviews relevant background and related work. The study design is presented in Section~\ref{sec:research}. Section~\ref{sec:results} outlines the results of the study, which are analyzed and discussed in Section~\ref{sec:discussion}. Finally, the conclusions are drawn in Section~\ref{sec:conclusion}.

\section{Background and Related Work}
\label{sec:background}

\subsection{Software Explainability} \label{sec:software-explainability}
Software explainability has become an increasingly popular research field in recent years, especially in the field of requirements engineering~\cite{chazette2021exploring,chazette2022can,chazette2020explainability,kohl2019explainability}.

One challenge in this field is the high degree of individuality in users' needs for explanation~\cite{ramos2021modeling}. In addition, different stakeholders (e.g., engineers, end-users, legal professionals) require different types of explanations~\cite{kohl2019explainability}. This variability makes it essential to correctly identify explanation requirements to ensure that user needs are met. In particular, it is important not to overestimate users' needs, as explanations may also negatively affect other non-functional requirements~\cite{chazette2021exploring}. For example, Chazette et al.~\cite{chazette2020explainability} revealed that explainability can impair a system's usability. Furthermore, explanations can negatively impact user experience~\cite{deters2024UXandExplainability}, causing stress~\cite{gruning2024stressful} and increasing users' cognitive load~\cite{nunes2017systematic}.
Chazette et al.~\cite{chazette2020explainability} proposed user-centered design techniques to mitigate these negative effects and recommended balancing the costs and benefits of providing explanations. Köhl et al.~\cite{kohl2019explainability} suggested treating explainability requirements as trade-offs during development.

To facilitate the handling of explainability requirements, several taxonomies have been created in previous work~\cite{unterbusch2023explanation, droste2024explanations}. Unterbusch et al.~\cite{unterbusch2023explanation} developed a taxonomy for explanation needs derived from app reviews, distinguishing between primary and secondary concerns. Droste et al.~\cite{droste2024explanations} created a taxonomy for explanation needs in everyday software systems using an online survey. They found that the categories \textit{interaction} and \textit{system behavior} were the most prevalent explanation needs in everyday software. Speith~\cite{speith2022XAITaxonomies} reviewed eleven existing explainability taxonomies and provided recommendations for their development.

Chazette et al.~\cite{chazette2022can} identified six key activities for developing explainable systems, which they validated through interviews with 19 practitioners. They emphasized that existing user-centered practices could be effectively utilized to gather and implement explainability requirements.

\subsection{Requirement Elicitation}

\subsubsection{Elicitation of Explainability Requirements}
\label{sec:elicitation-of-explainability-requirements}
In recent years, the elicitation of explainability requirements has gained increasing attention~\cite{droste2024explanations, unterbusch2023explanation,obaidi2025appKonwledge,obaidi2025mood,obaidi2025automatingexplanationneedmanagement,deters2024qualitymodel,obaidi2025explainability,obaidi2025AppFeaturesExplainNeeds,Deters2025quality,deters2025identifying}. Identifying the need for explanations is challenging because directly asking users what explanations they require may introduce hypothetical bias~\cite{hypotheticalBiasPlott} and the so-called \textit{why-not mentality}~\cite{droste2023designing}. The \textit{why-not mentality} refers to users' tendency to always answer affirmatively when asked whether they need an explanation, as they do not perceive any negative consequences. To mitigate these biases, Deters et al.~\cite{deters2024pulse} attempted to objectively detect the need for explanations using biometric data. However, their results showed that biometric data is not yet a reliable predictor of explanation needs.

Ramos et al.~\cite{ramos2021modeling} sought to support the elicitation process by developing explainability personas. Based on the responses of 61 users to a questionnaire, they identified five distinct personas to represent different explanation needs.

To systematically categorize explanation needs, Droste et al.~\cite{droste2024explanations} conducted an online survey with 84 participants, identifying explanation needs in everyday software systems. To enable structured categorization, they developed a taxonomy based on their study results, comprising five main categories: \textit{interaction}, \textit{system behavior}, \textit{privacy and security}, \textit{domain knowledge}, and \textit{user interface}.

To initiate a structured requirements engineering process for explainability, Chazette et al.~\cite{chazette2022framework} developed a quality framework for explainability, summarizing external dependencies, characteristics of explanations, and evaluation methods. Their framework facilitates the analysis, operationalization, and assessment of explainability requirements and was validated through a case study involving a navigation app.

\subsubsection{Requirement Elicitation in General}
The elicitation of requirements is a fundamental phase in requirements engineering, as it lays the foundation for all subsequent development activities~\cite{pacheco2018requirements}. The integration of stakeholders is crucial at this stage, as the software must be designed according to their needs and expectations~\cite{mishra2008successful}. Various methods exist for requirement elicitation, including interviews, surveys, observations, focus groups, brainstorming, and prototyping~\cite{younas2017non, alflen2020model, pacheco2018requirements}.

For complex projects with frequently changing requirements, Mishra et al.~\cite{mishra2008successful} recommend combining interviews, workshops, and iterative development to improve requirement accuracy and completeness.

Hadar et al.~\cite{hadar2014role} investigated the impact of domain knowledge on requirement elicitation through interviews. Their study revealed that domain knowledge can have both positive and negative effects on communication and the understanding of stakeholder needs.

\subsubsection{Comparison of Elicitation Methods}
\label{sec:anforderungserhebung-qualitative-oder-quantitative-erhebungsmethoden}
The selection and combination of appropriate elicitation techniques can significantly improve the quality and accuracy of collected requirements~\cite{alflen2020model}. Anwar and Razali~\cite{anwar2012practical} conducted an empirical study to establish practical guidelines for selecting requirement elicitation methods. They identified four main factors influencing the choice of methods: technical characteristics, stakeholder traits, requirement sources, and the project environment. Their study found that experts prefer conversational methods, such as interviews and workshops, when users have deep domain knowledge~\cite{anwar2012practical}. Conversely, questionnaires are more suitable when analysts possess some knowledge of the system, as they can guide users through the questions more effectively~\cite{anwar2012practical}.

In 2015, Pacheco et al.~\cite{pacheco2018requirements} conducted a systematic literature review on frequently used elicitation methods. They found that approximately 22\% of the studies employed more than one method, suggesting that combining multiple approaches is beneficial. However, they did not provide specific recommendations on how to integrate these methods, noting that each has unique advantages in different situations.

Younas et al.~\cite{younas2017non} highlighted the need to address non-functional requirements (NFRs) early in the software development process, as they influence technology selection, hardware allocation, and security standards.

\subsection{Selection of Elicitation Methods}
Requirements engineering research indicates that combining multiple elicitation methods can be effective~\cite{alflen2021using,jiang2005combining,sabahat2010iterative}. In this context, conversational techniques such as interviews are among the most commonly used~\cite{alflen2020model,anwar2012practical}. Workshops, focus groups, and interviews are all considered conversational methods. Interviews are particularly effective when the system involves users with different roles~\cite{anwar2012practical} and are beneficial in both global software development and traditional environments for gathering detailed information~\cite{pacheco2018requirements}. They are preferred because they foster deeper discussions through communication~\cite{anwar2012practical}.
Workshops help resolve discrepancies between users~\cite{anwar2012practical}, encouraging collaboration and discussion~\cite{pacheco2018requirements}, and are often recommended for requirement elicitation~\cite{tiwari2017methodology}. A study by Chazette et al.~\cite{chazette2022can} found that interviews, focus groups, workshops, surveys, and personas are among the most effective methods for eliciting explainability requirements, as reported by experienced IT professionals.
Krüger~\cite{krueger2014focus} defines focus groups as collaborative discussions aimed at elaborating stakeholder opinions, with a moderator facilitating a comfortable environment for discussion. Focus groups are effective for gathering multiple opinions to formalize requirements~\cite{pacheco2018requirements}. Given the aim of this work to identify as many explainability needs as possible, focus groups were selected, as users require an introduction to explainability, and expert moderators can guide discussions effectively. 
For large user groups, surveys are recommended~\cite{anwar2012practical}, although they require careful planning of questions~\cite{anwar2012practical}. However, surveys alone are less suitable for analyzing user experience~\cite{fehlmann2013customer}, and should be combined with other methods~\cite{carod2010cognitive,droste2023designing}. Interviews, as the primary method, enhance requirement quality when combined with other techniques~\cite{sabahat2010iterative}. 

\section{Study Design}
\label{sec:research}
We conducted a comparative analysis of three requirement elicitation methods -- focus groups, interviews, and an online survey -- within a company that uses personnel management software. Our objective was to determine which elicitation method is the most effective and efficient for capturing explainability requirements.  

The design of the overall research for this work is illustrated in Figure~\ref{fig:online-studie}. Our process began with a requirements engineer designing a structured elicitation procedure, which served as the foundation for all three elicitation methods. This included the application of an existing taxonomy for categorizing explanation needs into categories such as \textit{Interaction} or \textit{System behavior}. Based on the results of the focus groups and interviews, the online survey was conducted using the delayed taxonomy approach. Once the survey was completed, a third categorized list of explanation needs was generated. The lists of needs collected throughout the three elicitation methods were then consolidated, analyzed, and evaluated to address the research questions. Eventually, the evaluation resulted in a comprehensive coded list of explanation needs. 

\begin{figure*}[htbp]
    \centering
    \includegraphics[width=0.8\linewidth]{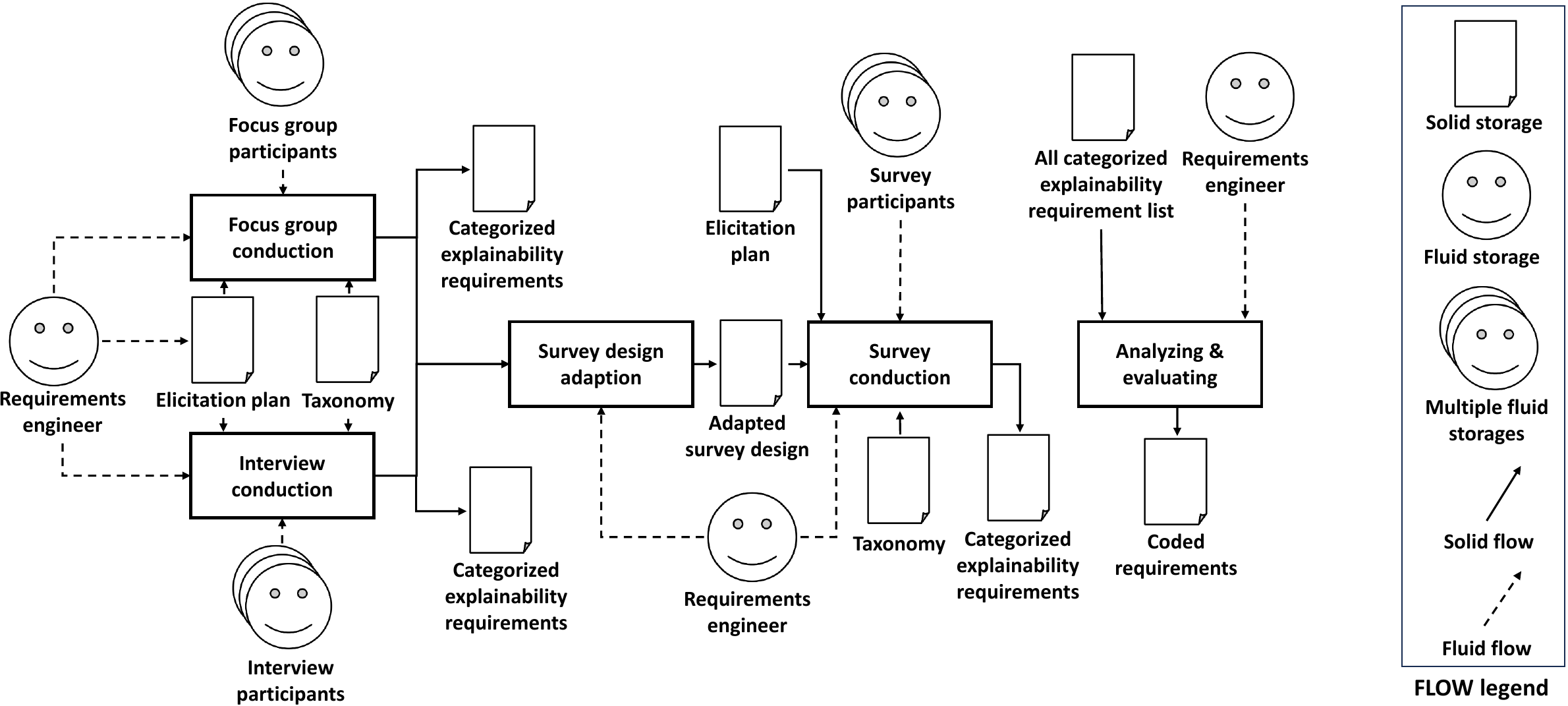}
    \caption{Overview of our research design in FLOW notation~\cite{stapel2009flow}.}
    \label{fig:online-studie}
\end{figure*}

\subsection{Research Goal and Research Questions}

We strive to achieve the following goal, formulated according to Wohlin et al.'s Goal-Definition Template~\cite{wohlin2012experimentation}:\\

\setlength{\shadowsize}{2pt}
\noindent
\shadowbox{
\begin{minipage}[t]{0.95\columnwidth}
\textbf{Research Goal:} \textit{Compare} the efficiency and effectiveness of different elicitation methods
\textit{for the purpose of} identifying the most suitable approach for gathering explanation needs
\textit{with respect to} the elicitation methods focus groups, interviews, and online surveys
\textit{from the point of view of} a requirements engineer
\textit{in the context of} explainability requirements elicitation.
\end{minipage}
}\vspace{0pt}

We investigate the following research questions:

\begin{itemize}

\item \textbf{RQ1: Which of the elicitation methods (focus groups, interviews, or online surveys) is the most \textit{efficient} for collecting explainability requirements?}
Answering this research question will help us identify which data collection method optimizes resource usage, such as time and effort, while collecting explainability requirements from users.

\item \textbf{RQ2: Which of the elicitation methods (focus groups, interviews, or online surveys) is the most \textit{effective} for collecting explainability requirements?}
This question will allow us to determine which method generates the highest quality and most comprehensive data regarding explainability requirements, helping to guide future studies in this area.

\item \textbf{RQ3: How do the \textit{results} from the focus groups, interviews, and online survey \textit{differ}?}
By addressing this question, we can compare how different methods influence the types and frequencies of explanation needs captured, providing insight into the strengths and limitations of each method.

\item \textbf{RQ4: How do the collected explanation needs differ depending on when an explainability need taxonomy is applied?}
The answer to this question will help us understand the impact of introducing a taxonomy at different stages of data collection on the comprehensiveness and categorization of explanation needs, providing guidance on when to apply a taxonomy to maximize the quality and completeness of elicited requirements in practice.

\end{itemize}

\subsection{Participant Recruitment}
Participants were recruited from a large German IT consulting company that actively uses the personnel management software examined in this study. The focus groups were conducted in person at company offices, while interviews and the online survey were administered remotely. The survey was distributed using LimeSurvey, ensuring broad accessibility for employees across different locations.

\subsection{Methodology to Compare the Different Elicitation Methods}
This paper compares three different elicitation methods. To support the elicitation processes, we provided the stakeholders with a taxonomy that details possible types of explanation needs. In particular, we used the taxonomy by Droste et al.~\cite{droste2024explanations}, which serves as a checklist and provides a guideline for the requirements engineer to identify the desired explanation needs. 

To enable a detailed analysis, we designed two main study variants: one with direct taxonomy usage from the outset, and one where the taxonomy was introduced only after an initial open elicitation phase. From these two variants, we derived and analyzed three conditions: \textit{without taxonomy} (before introduction), \textit{direct taxonomy usage}, and \textit{delayed taxonomy usage} (after initial open elicitation followed by taxonomy introduction).

For both focus groups and interviews, these two variants were implemented identically: In the first variant, the taxonomy was used from the start (“direct taxonomy usage”). In the second, needs were first collected openly without the taxonomy, after which the taxonomy was introduced to gather additional requirements (“delayed taxonomy usage”). This design allowed us to compare the \textit{without} and \textit{delayed} taxonomy conditions within the same group of participants.


For the online survey, the same process as used in the focus groups and interviews served as the basis for its design. Insights from the previously conducted qualitative methods guided the decision to apply the delayed taxonomy usage variant in the survey, as it yielded the highest number of distinct needs while maintaining efficiency.

Tables \ref{tab:vorgehen} and \ref{tab:vorgehen2} provide an overview of the steps in the two different study procedures. In the first approach, the taxonomy was introduced at the beginning and used immediately to elicit explainability requirements (\textit{direct} condition). In the second approach, requirements were initially collected without the taxonomy (\textit{without} condition), allowing participants to express their needs freely. The taxonomy was then introduced, and additional requirements were gathered from the same participants, enabling the analysis of the \textit{delayed} taxonomy usage condition. In this version, questions that were previously asked at the end of the session as closing questions were incorporated as interim questions to enhance the elicitation process.

\begin{table}[htb]
\scriptsize
\caption{Overview of the steps of the methods with details, for the version with direct taxonomy introduction.}
\label{tab:vorgehen}
\centering
\begin{tabularx}{\columnwidth}{rp{3cm}X}
\toprule
& \textbf{Step} & \textbf{Details} \\ \midrule
\textbf{1}   & Introduction & Greeting + introduction, Purpose + duration of the survey \\ \midrule
\textbf{2}   & Background Information & Role, Gender + age \\ \midrule
\textbf{3}   & General Usage Information & Duration of use, Frequency of use \\ \midrule
\textbf{4}   & Software Description and Use & Have the software introduced, Allow 3 minutes for software use \\ \midrule
\textbf{5}   & Introduction to Explainability & Brief introduction to the topic, Introduction to explanation needs \\ \midrule
\textbf{6}   & Introduction to Taxonomy with Examples & Going through the taxonomy, Associated example list \\ \midrule
\textbf{7}   & Identifying Explanation Needs & Collecting needs using the taxonomy \\ \midrule
\textbf{8}   & Categorization of Needs & Categorizing the mentioned needs using the taxonomy \\ \midrule
\textbf{9}   & Closing Questions & Ending the survey with a follow-up question \\ \bottomrule
\end{tabularx}
\end{table}

\begin{table}[htb]
\scriptsize
\caption{Overview of the steps of the methods with details, for the version with later taxonomy introduction with changes marked compared to the direct taxonomy introduction version.}
\label{tab:vorgehen2}
\begin{tabularx}{\columnwidth}{rXX}
\toprule
& \textbf{Step} & \textbf{Details} \\ \midrule
\textbf{5} & Introduction to Explainability & Brief introduction to the topic, ... \\ \midrule
\textbf{6} & \ul{Identifying Explanation Needs without Taxonomy} & \ul{Collecting the needs} \\ \midrule
\textbf{7} & \ul{Interim Questions} & \ul{Answering interim questions} \\ \midrule
\textbf{8} & Introduction to Taxonomy with Examples & Going through the taxonomy, ... \\ \bottomrule
\end{tabularx}
\end{table}


\subsubsection{Focus Groups}

Two focus groups were conducted, each consisting of six participants. One group elicited explanation needs using the taxonomy from the beginning, while the other group first collected requirements without the taxonomy and was then introduced to it afterward. Each session lasted between 71 and 73 minutes.

\subsubsection{Interviews}

A total of 18 participants took part in the interviews, with nine assigned to the direct taxonomy usage group and nine to the group without an initial taxonomy usage. All interviews were conducted online.
For the group with immediate taxonomy usage, the interviews lasted between 6 and 20 minutes, with an average duration of 11:53 minutes.
For the group without an initial taxonomy usage, the duration ranged from 5 to 22 minutes, with an average of 11:07 minutes.
In the group where requirements were first elicited without the taxonomy and later with its usage, the interviews lasted between 8 and 33 minutes, with an average duration of 15:53 minutes.

\subsubsection{Survey}

For the online survey, the variant with delayed taxonomy usage was chosen. This decision was based on the increased average number of explanation needs per participant, as indicated in Table \ref{table:Overview_Methods}. The survey was designed and conducted using LimeSurvey\footnote{\url{https://www.limesurvey.org/}}.
A total of 895 participants started the survey, of whom 277 completed it. However, responses containing irrelevant answers, such as \enquote{I have no needs} or nonsensical input (e.g., \enquote{??????}), were excluded from the final dataset. After filtering, 188 valid responses remained for analysis.
The time participants spent completing the survey varied significantly. For participants who answered without taxonomy usage, completion times ranged from 1:54 minutes to 62:22 minutes, with an average of 11:24 minutes and a median completion time of 9:16 minutes.
For participants who first provided responses without the taxonomy and then continued with its usage, completion times ranged from 2:28 minutes to 87:28 minutes, with an average duration of 14:08 minutes and a median of 11:09 minutes.

\subsection{Data Analysis}

\subsubsection{Metrics}
To compare the efficiency of the elicitation methods, the time required to collect the explanation needs was measured. The total study time represents the entire duration from the beginning to the end of each focus group, interview, or survey. This includes both the moderator, who conducts the focus groups and interviews, and all study participants. The total effort required for each method is calculated by multiplying the total study time by the number of participants involved:
\begin{equation*}
\text{Personal effort} = \text{Total study time} \cdot \text{Number of participants}
\end{equation*}
The elicitation time consists of the time spent on introducing the study to the participants and the actual process of collecting explanation needs. Participants also received an introduction to the concept of explainability before identifying their explanation needs, which was included in the calculation.

To assess the effectiveness of the elicitation methods, the absolute number of explanation needs collected was analyzed. The total number of explanation needs was then divided by the number of participants to account for differences in sample sizes, providing a normalized measure of effectiveness.
Additionally, the distribution of needs across different categories, as defined by the taxonomy of Droste et al.~\cite{droste2024explanations}, was examined to gain insights into the ability of each method to capture a diverse range of explanation needs.

\subsubsection{Types of Explanation Needs}
To evaluate the effectiveness of the different elicitation methods, all explanation needs were categorized based on the taxonomy by Droste et al.~\cite{droste2024explanations}. Although somewhat recent, the taxonomy has already been applied in practice by other researchers~\cite{obaidi2025automatingexplanationneedmanagement,obaidi2025mood,Droste2025REJexpl}, which underlines its value in practice. The categorization was performed by one requirements engineer with experience in explainability. In cases of uncertainty, a second requirements engineer, also experienced in explainability, was consulted to ensure a consistent and accurate classification. Furthermore, the taxonomy was extended in alignment with Obaidi et al.~\cite{obaidi2025automatingexplanationneedmanagement} and supplemented with additional software-specific categories to allow the identification of multiple explanation needs per response. This extension enables the determination of the distinct number of explanation needs.

\section{Results}
\label{sec:results}

The anonymized dataset, including the questionnaires, is available on \href{https://doi.org/10.5281/zenodo.15678937}{Zenodo}~\cite{obaidi2025datasetelicit}. The quantitative results of the elicitation methods are summarized in Table \ref{table:Overview_Methods}.

\begin{table*}[htb]
\scriptsize
\caption{Statistical comparison of the elicitation methods, distinguishing the versions of taxonomy usage, with the best values per aspect marked.}
\label{table:Overview_Methods}
\begin{tabularx}{\textwidth}{X | r r r | r r r | r r}
\toprule
\textbf{} & \multicolumn{3}{c}{Focus group} & \multicolumn{3}{|c}{Interviews} & \multicolumn{2}{|c}{Survey} \\
\midrule
\textbf{Taxonomy usage} & None & Direct & Delayed & None & Direct & Delayed & None & Delayed \\
\midrule
\textbf{\# Participants} & 6 & 6 & 6 & 9 & 9 & 9 & 188 & 188 \\
\textbf{\# Needs} & 20 & 27 & 28 & 103 & \textbf{109} & 147 & \textbf{409} & \textbf{471} \\
\textbf{\# Distinct needs} & 19 & 23 & 27 & 96 & \textbf{105} & 133 & \textbf{327} & \textbf{364} \\
\midrule
\textbf{Personnel effort [h]} & \textbf{3:13} & \textbf{6:20} & \textbf{6:04} & 7:34 & 10:21 & 11:35 & 35:25 & 39:01 \\
\textbf{Average total time [min]} & 27:36 & 54:18 & 52:00 & 23:28 & \textbf{33:59} & 38:13 & \textbf{11:24} & \textbf{14:08} \\
\textbf{Average elicitation time [min]} & 12:30 & \textbf{18:48} & 25:54 & 10:40 & 20:18 & 20:20 & \textbf{4:06} & \textbf{6:50} \\
\textbf{Average distinct needs per participant} & 3.17 & 3.83 & 4.50 & \textbf{10.67} & \textbf{11.67} & \textbf{14.78} & 1.74 & 1.94 \\
\textbf{Distinct needs per participant per average time} & 0.15 & 0.12 & 0.10 & \textbf{0.66} & \textbf{0.34} & \textbf{0.43} & 0.25 & 0.17 \\
\textbf{Distinct needs per personnel effort} & 5.54 & 3.38 & 4.27 & \textbf{12.42} & \textbf{10.08} & \textbf{11.29} & 9.14 & 9.20 \\
\bottomrule
\end{tabularx}
\end{table*}

The survey method, particularly with delayed taxonomy usage, yielded the highest number of total (471) and distinct needs (364). However, interviews with delayed taxonomy usage produced the highest number of distinct needs per participant (14.78), suggesting its effectiveness in capturing diverse explanation needs. 
Regarding efficiency, focus groups and interviews required significantly less personnel effort compared to the survey. The shortest average total and elicitation times were recorded for the survey with direct taxonomy usage (11:24 min and 4:06 min, respectively). However, when normalizing for participant effort, interviews with direct taxonomy usage demonstrated the highest number of distinct needs per personnel hour (12.42). 
Overall, the results indicate that while surveys allow for broad data collection, interviews—particularly with delayed taxonomy usage—offer a more efficient and focused means of capturing diverse and distinct explanation needs.

Table \ref{table:Elicitation_Comparison_Full} provides an overview of the number of distinct needs per taxonomy category.

\begin{table*}[htb]
\scriptsize
\caption{Comparison of elicitation methods and taxonomy usage across absolute values and percentage distributions. The highest numbers are highlighted.}
\label{table:Elicitation_Comparison_Full}
\begin{tabularx}{\textwidth}{r l l r r r r r r r r}
\toprule
\textbf{\# Participants} & \textbf{Elicit. Method} & \textbf{Tax. Usage} & \textbf{System behavior} & \textbf{Interaction} & \textbf{Feature missing} & \textbf{Domain} & \textbf{Business} & \textbf{Sec. \& priv.} & \textbf{UI} & \textbf{Total} \\
\midrule
6   & Focus Group   & Without   & 3 (16\%)  & 4 (\textbf{21\%})  & 4 (\textbf{21\%})  & 5 (26\%)  & 0 (0\%)  & \textbf{1} (\textbf{5\%})  & 2 (11\%)  & 19  \\
9   & Interviews    & Without   & 22 (\textbf{23\%})  & 9 (9\%)   & 12 (13\%)  & 41 (\textbf{43\%})  & 0 (0\%)  & 0 (0\%)  & 12 (13\%)  & 96  \\
188 & Survey       & Without   & \textbf{70} (21\%)  & \textbf{47} (14\%)  & \textbf{49} (15\%)  & \textbf{95} (29\%)  & \textbf{6} (\textbf{2\%})   & \textbf{1} (0\%)  & \textbf{59} (\textbf{18\%)}  & \textbf{327}  \\
\midrule
6   & Focus Group   & Direct    & 4 (17\%)  & 1 (4\%)   & 1 (4\%)   & 13 (\textbf{57\%})  & 0 (0\%)  & 1 (\textbf{4\%})  & 3 (13\%)  & 23  \\
9   & Interviews    & Direct    & \textbf{27} (\textbf{26\%})  & \textbf{10} (\textbf{10\%)}   & \textbf{7} (\textbf{7\%})   & \textbf{42} (40\%)  & 0 (0\%)  & \textbf{3} (3\%)  & \textbf{16} (\textbf{15\%})  & \textbf{105}  \\
\midrule
6   & Focus Group   & Delayed   & 4 (15\%)  & 4 (\textbf{15\%})  & 6 (\textbf{22\%})  & 8 (30\%)  & 0 (0\%)  & 2 (\textbf{7\%})  & 3 (11\%)  & 27  \\
9   & Interviews    & Delayed   & 27 (20\%)  & 14 (11\%)  & 16 (12\%)  & 54 (\textbf{41\%})  & 0 (0\%)  & 3 (2\%)  & 19 (14\%)  & 133  \\
188 & Survey       & Delayed   & \textbf{77} (\textbf{21\%})  & \textbf{52} (14\%)  & \textbf{58} (16\%)  & \textbf{102} (28\%)  & \textbf{9} (\textbf{2\%})   & \textbf{4} (1\%)  & \textbf{62} (\textbf{17\%})  & \textbf{364}  \\
\bottomrule
\end{tabularx}
\end{table*}

The results show that the survey method collected the highest number of distinct needs (364), followed by interviews (133) and focus groups (27). However, the percentage distribution of explanation needs across taxonomy categories varies between methods. While focus groups resulted in a higher proportion of needs related to \textit{Feature missing} (up to 22\%) and \textit{Domain} (26–57\%), interviews captured a relatively balanced distribution across categories, with a notable emphasis on \textit{System behavior} (20–26\%) and \textit{Feature missing} (7–13\%). Surveys, in contrast, had the highest share of \textit{UI} needs (up to 18\%), suggesting that larger-scale participation may surface additional concerns in this category.

Figure \ref{fig:evaluation-without} presents the number of distinct explanation needs per participant, categorized by explanation need type, without the use of a taxonomy. The normalized values reveal notable differences between elicitation methods. Interviews generated the highest number of needs per participant (10.7), significantly outperforming focus groups (3.2) and surveys (1.7). Across categories, interviews elicited a particularly high number of \textit{Domain} needs (4.6 per participant), while focus groups distributed explanation needs more evenly. Surveys resulted in relatively fewer needs per category, indicating that large-scale data collection may yield a broader but less detailed set of needs per respondent.

\begin{figure}[htb]
    \centering
    \includegraphics[width=1\linewidth]{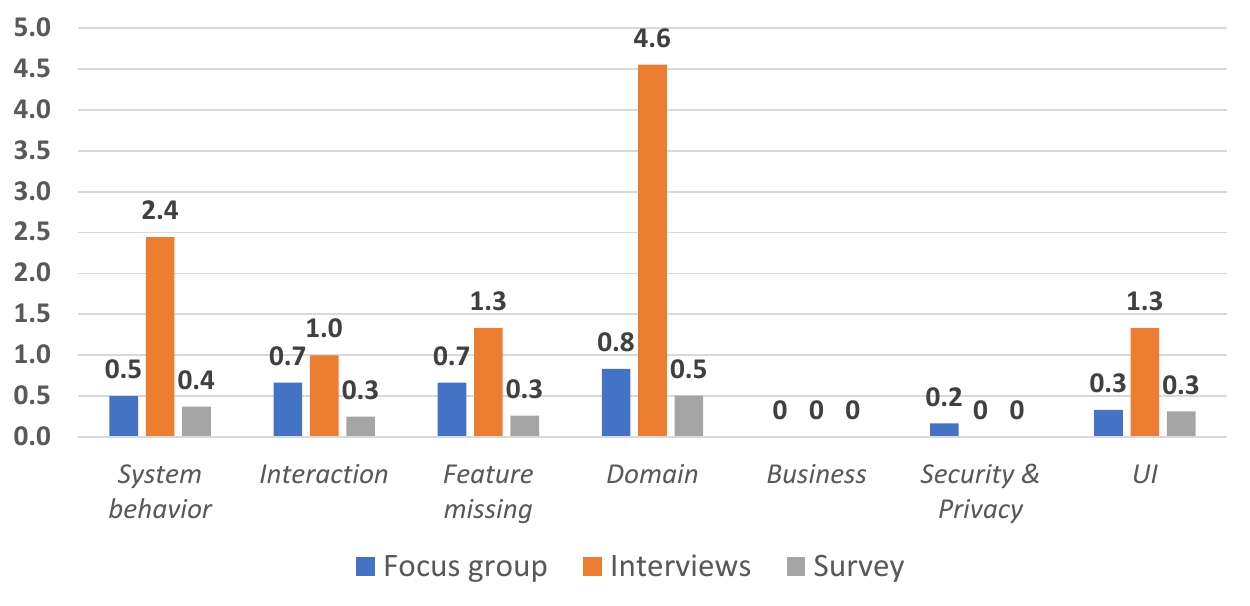}
    \caption{Number of explanation needs per participant without taxonomy usage, categorized by explanation need type.}
    \label{fig:evaluation-without}
\end{figure}

Figure \ref{fig:evaluation-direct} presents the number of distinct explanation needs per participant when the taxonomy was introduced at the beginning. The results show that interviews yielded the highest number of explanation needs per participant (11.7), significantly outperforming focus groups (3.8). This suggests that the structured nature of interviews, where individuals express their needs without group influence, may facilitate a broader collection of distinct explanation needs. In contrast, focus groups, while still effective, resulted in a lower number of needs per participant, indicating that group discussions may lead to shared perspectives and reduced individual variance. Across both methods, the domain-related category exhibited the highest number of needs, particularly in interviews (4.7 per participant), reinforcing the idea that taxonomy guidance helps participants articulate their domain-specific explanation needs.

\begin{figure}[htb]
    \centering
    \includegraphics[width=1\linewidth]{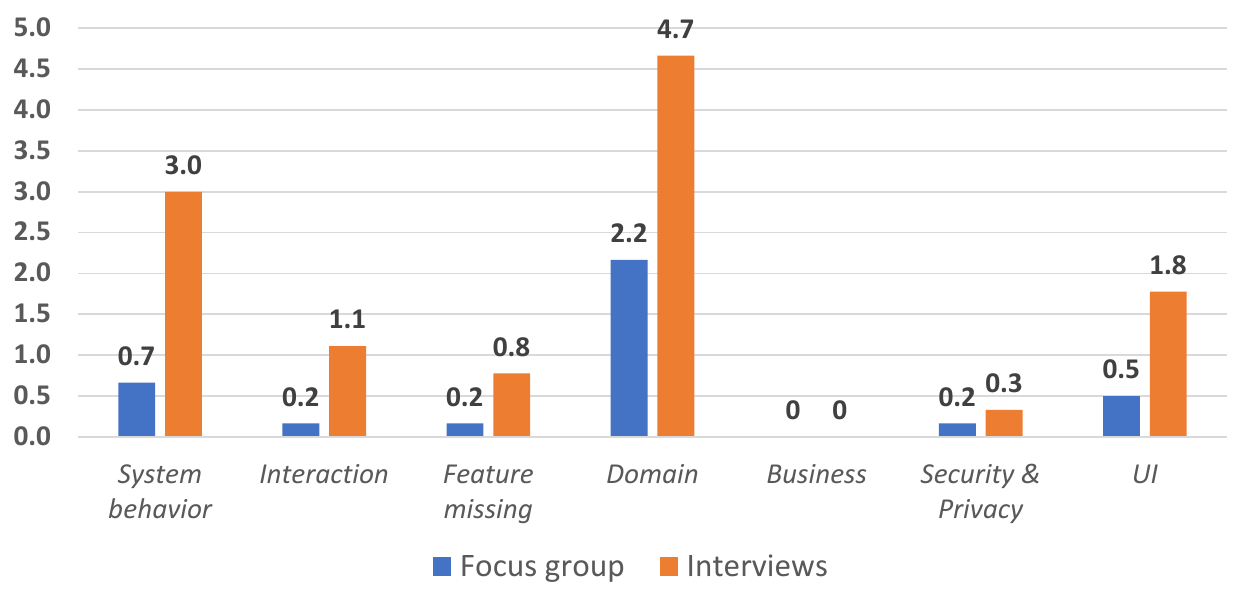}
    \caption{Number of explanation needs per participant with direct taxonomy usage, categorized by explanation need type.}
    \label{fig:evaluation-direct}
\end{figure}

Figure \ref{fig:evaluation-delayed} illustrates the number of distinct explanation needs per participant when the taxonomy was introduced after an initial phase without it. As seen in the direct taxonomy usage scenario, interviews again resulted in the highest number of needs per participant (14.8), followed by focus groups (4.5) and surveys (1.9). The increase in collected needs compared to direct taxonomy usage (especially in interviews) suggests that allowing participants to express their needs freely before introducing a structured taxonomy may help them articulate a broader range of requirements. In both focus groups and interviews, domain-related needs remained the most frequently mentioned category, with 6.0 per participant in interviews and 1.3 in focus groups, highlighting the importance of domain-specific explainability concerns. The survey method, while having the lowest per-participant count, still showed a slight increase compared to its direct taxonomy usage counterpart, suggesting that a delayed taxonomy introduction may provide benefits in guiding participants without restricting their initial thought process.

\begin{figure}[htb]
    \centering
    \includegraphics[width=1\linewidth]{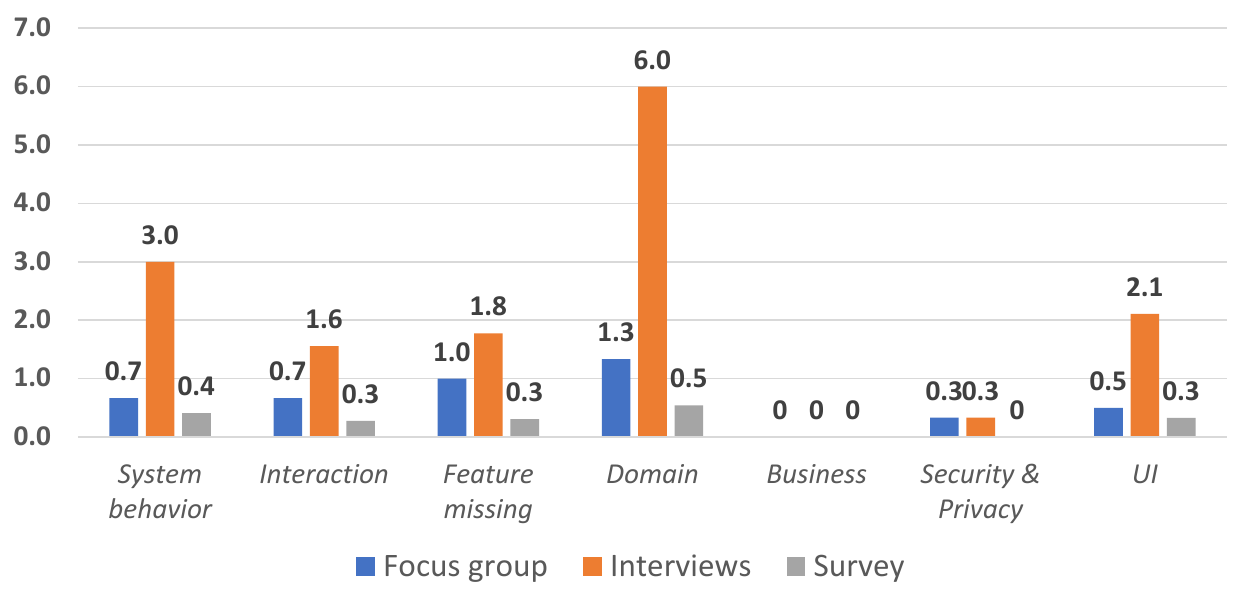}
    \caption{Number of explanation needs per participant with delayed taxonomy usage, categorized by explanation need type.}
    \label{fig:evaluation-delayed}
\end{figure}

\begin{table*}[htb]
\scriptsize
\caption{Comparison of elicitation methods and taxonomy usage, showing the proportion of total needs to distinct needs, with category-wise distributions. The highest values per taxonomy usage group are bolded.}
\label{table:Elicitation_Comparison}
\begin{tabularx}{\textwidth}{l l r | r r r r r r r | r}
\toprule
\textbf{Elicit. Method} & \textbf{Tax. Usage} & \textbf{Needs / Distinct} & \textbf{System behavior} & \textbf{Interaction} & \textbf{Feature missing} & \textbf{Domain} & \textbf{Business} & \textbf{Sec.} \& \textbf{Priv.} & \textbf{UI} & \textbf{Total} \\
\midrule
Focus Group   & Without   & 20 / 19   & 0.00\%  & 0.00\%  & 0.00\%  & \textbf{16.67\%}  & 0.00\%  & 0.00\%  & 0.00\%  & 5.00\%  \\
Focus Group   & Direct    & \textbf{27 / 23}   & 0.00\%  & 0.00\%  & 0.00\%  & \textbf{18.75\%}  & 0.00\%  & \textbf{50.00\%}  & 0.00\%  & \textbf{14.81\%}  \\
Focus Group   & Delayed   & 28 / 27   & 0.00\%  & 0.00\%  & 0.00\%  & 11.11\%  & 0.00\%  & 0.00\%  & 0.00\%  & 3.57\%  \\
\midrule
Interviews    & Without   & 103 / 96  & 4.35\%  & 0.00\%  & \textbf{7.69\%}  & \textbf{10.87\%}  & 0.00\%  & 0.00\%  & 0.00\%  & 6.80\%  \\
Interviews    & Direct    & 109 / 105 & 0.00\%  & \textbf{9.09\%}  & 0.00\%  & 6.67\%   & 0.00\%  & 0.00\%  & 0.00\%  & 3.67\%  \\
Interviews    & Delayed   & 147 / 133 & 6.90\%  & 6.67\%  & \textbf{15.79\%} & 12.90\%  & 0.00\%  & 0.00\%  & 0.00\%  & 9.52\%  \\
\midrule
Survey        & Without   & \textbf{409 / 327} & \textbf{17.65\%} & 9.62\%  & 3.92\%  & 34.48\%  & 0.00\%  & 0.00\%  & \textbf{14.49\%} & \textbf{20.05\%} \\
Survey        & Delayed   & \textbf{471 / 364} & \textbf{21.43\%} & \textbf{11.86\%} & 6.45\%  & \textbf{36.65\%}  & 0.00\%  & \textbf{20.00\%}  & \textbf{19.48\%} & \textbf{22.72\%} \\
\bottomrule
\end{tabularx}
\end{table*}

The comparison of taxonomy usage across elicitation methods reveals that delayed taxonomy usage consistently led to the highest number of explanation needs per participant. This effect was particularly strong in interviews, where the delayed approach resulted in 14.8 needs per participant, compared to 11.7 with direct usage and 10.7 without taxonomy. The increase was especially notable for domain-related and feature-missing needs, suggesting that an initial open-ended elicitation phase helps participants articulate more detailed requirements before being guided by the taxonomy. While focus groups showed a smaller increase, surveys exhibited only minor differences, indicating that structured input formats benefit less from delayed taxonomy introduction than interactive methods like interviews and discussions.

Table \ref{table:Elicitation_Comparison} highlights that the survey resulted in the highest number of repeated explanation needs, which is expected given the large number of participants (188 after filtering). The proportion of explanation needs mentioned multiple times was 20.05\% without taxonomy usage and 22.72\% with taxonomy usage, with the highest individual explanation need being repeated 17 times. Interestingly, focus groups exhibited the highest proportion of repeated needs when the taxonomy was introduced directly (14.81\%), suggesting that participants may have been more guided in their responses. In comparison, interviews showed a much lower proportion of repeated needs, with only 3.67\% under direct taxonomy usage. This indicates that interactive settings, particularly those with direct moderation, lead to a broader distribution of explanation needs, while surveys facilitate higher redundancy in the collected requirements.

\begin{figure}[htb]
    \centering
    \subfloat[Without taxonomy]{%
        \includegraphics[width=0.4\linewidth]{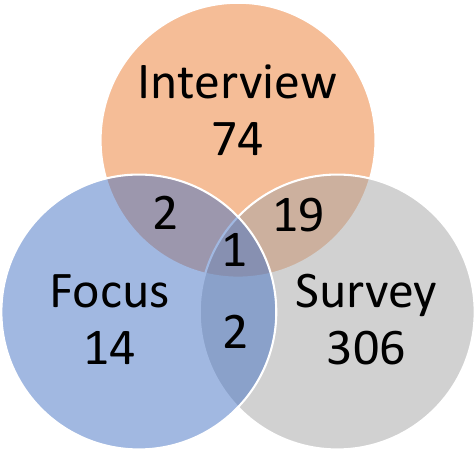}
        \label{fig:venn-without}
    }
    \hfill
    \subfloat[Delayed taxonomy usage]{%
        \includegraphics[width=0.4\linewidth]{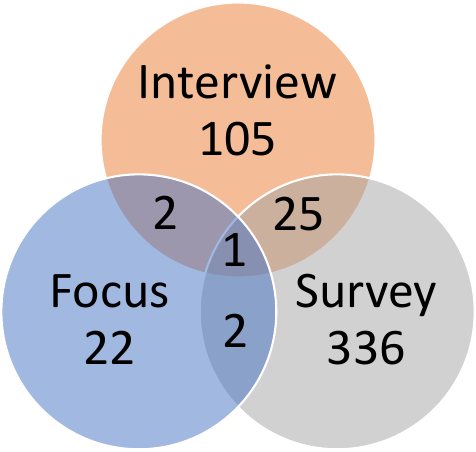}
        \label{fig:venn-delayed}
    }
    \caption{Comparison of explanation needs across elicitation methods: left without taxonomy usage, right with delayed taxonomy usage.}
    \label{fig:venn-comparison}
\end{figure}

The Venn diagrams in Figure \ref{fig:venn-comparison} illustrate the overlap of explanation needs across elicitation methods for two different taxonomy usage conditions: without taxonomy usage and with delayed taxonomy usage.

For the condition without taxonomy usage (Figure \ref{fig:venn-without}), the majority of needs remain unique to each method, with 327 needs exclusive to the survey, 96 to interviews, and 19 to focus groups. Overlap between methods is limited, with 24 needs shared between surveys and interviews, and only one need identified across all three methods.

For the delayed taxonomy usage condition (Figure \ref{fig:venn-delayed}), the overall trend remains similar, with the majority of needs being method-specific. Here, 336 needs are unique to surveys, 105 to interviews, and 22 to focus groups, while 25 needs overlap between interviews and surveys. Notably, the introduction of the taxonomy after an initial round of elicitation does not significantly alter the overlap between methods, but it does lead to a slightly higher total number of identified needs.

For the direct taxonomy usage condition, only focus groups and interviews were conducted, with 20 needs exclusive to focus groups, 102 to interviews, and only three shared between both. This indicates that direct taxonomy usage does not necessarily increase the overlap between methods, but rather helps structure the responses within each method.

\section{Discussion}
\label{sec:discussion}

In the following, we answer the research questions, present threats to validity, and interpret the results.

\subsection{Answers to the Research Questions}
\label{sec:beantworten-der-forschungsfragen}

\textbf{RQ1: Which of the elicitation methods (focus groups, interviews, or online surveys) is the most efficient for collecting explainability requirements?}  
Interviews were the most efficient elicitation method, achieving the highest values for distinct needs per participant per average time (0.34 or higher) and per personnel effort (10.08 or higher), while surveys followed closely. Focus groups were the least efficient, with values below 0.15 and 5.54, respectively. However, when considering absolute numbers, surveys collected the highest total number of explanation needs, making them the most productive method in terms of sheer volume.

\textbf{RQ2: Which of the elicitation methods (focus groups, interviews, or online surveys) is the most effective for collecting explainability requirements?}  
Surveys collected the highest total number of explanation needs, but many were repeated, with 20.05\% to 22.72\% redundancy depending on taxonomy usage. Interviews had the highest number of distinct needs per participant, making them the most effective method for capturing a diverse set of explainability requirements.  

\textbf{RQ3: How do the \textit{results} from the focus groups, interviews, and online survey \textit{differ}?}  
The distribution of explanation needs varied across methods, with interviews and focus groups eliciting more domain-related and feature-missing needs, while surveys captured a broader spread across categories but with more repetition. Business-related needs were primarily captured in focus groups, while \textit{Security \& privacy} needs were almost exclusively found in surveys with delayed taxonomy usage. 

\textbf{RQ4: How do the collected explanation needs differ depending on when an explainability need taxonomy is applied?}  
Delayed taxonomy usage led to the highest number of distinct needs per participant, especially in interviews, where it increased the number of elicited needs by approximately 26\%. Direct taxonomy usage resulted in a more structured distribution across categories, while no taxonomy usage produced fewer needs overall, particularly for less intuitive categories like \textit{Security \& privacy}.

\subsection{Interpretation}
\label{sec:interpretation}

The results of our study provide several insights into the efficiency and effectiveness of different elicitation methods for explainability requirements. Interviews emerged as the most efficient method, achieving the highest values for distinct needs per participant per average time, while focus groups were the least efficient. However, surveys collected the highest number of total and distinct needs, making them the most effective in absolute terms. This aligns with prior research~\cite{anwar2012practical, pacheco2018requirements}, which has shown that interviews and surveys can yield large amounts of data, while workshops and focus groups offer more structured, interactive discussions.

Beyond efficiency and effectiveness, our findings reveal that different elicitation methods tend to capture distinct types of explanation needs. Interviews elicited a particularly high proportion of \textit{Domain} needs, while surveys identified more \textit{System behavior} and \textit{User Interface} needs. Focus groups, despite yielding fewer overall needs, showed a notable share of \textit{Feature missing} needs. The low overlap between methods further supports that they elicit different perspectives on explainability requirements, emphasizing the necessity of a mixed-methods approach to achieve comprehensive coverage.

\subsubsection{The Role of Taxonomy Usage} 
One of the key research questions in our study was whether the use of a taxonomy improves the elicitation of explainability requirements. The results indicate that delayed taxonomy usage yields the highest number of distinct needs, particularly in interviews and surveys. This suggests that initially allowing participants to articulate their explanation needs without a predefined structure encourages more spontaneous and diverse responses. Only afterward, when the taxonomy is introduced, do participants benefit from additional guidance, helping them refine and categorize their needs. This insight supports the findings of prior research~\cite{droste2024explanations}, which highlights the importance of structured taxonomies for requirements management. However, the results suggest that introducing a taxonomy too early may limit creativity and reduce the diversity of elicited needs.

Furthermore, our findings indicate that surveys had a high dropout rate, particularly after the software usage questions and the introduction to explainability. This suggests that participants may expect surveys to be quick and straightforward and may disengage when they perceive them as too complex or time-consuming. This is an important consideration for companies using surveys to elicit requirements, as overly technical or lengthy surveys may deter participation. A more concise and user-friendly survey design could improve completion rates while still capturing valuable insights.

\subsubsection{Recommendations for Companies}  
Based on these findings, we propose the following recommendations for companies seeking to elicit explainability requirements:
\begin{enumerate}
    \item Use interviews for efficiency and surveys for scalability. If time and resources are limited, interviews provide the best balance between depth and efficiency. If a broader range of needs is required, surveys should be preferred.
    \item Consider a two-phase elicitation approach. First, collect explanation needs openly without imposing a predefined taxonomy, then introduce a structured taxonomy to refine and categorize responses. This method maximizes creativity while ensuring completeness.  
    \item Combine multiple elicitation methods. Prior studies~\cite{pacheco2018requirements, alflen2020model} suggest that combining qualitative and quantitative methods leads to better requirements coverage. Our findings reinforce this by demonstrating that surveys, interviews, and focus groups all have unique strengths.
    \item Ensure active moderation in focus groups. Since group discussions do not inherently prevent redundant responses, facilitators should encourage participants to build on previous answers rather than repeat them.
\end{enumerate} 

\subsubsection{Contributions} 
Our study builds on existing literature by providing an empirical comparison of elicitation methods specifically for explainability requirements. While previous studies have compared requirement elicitation methods in general~\cite{anwar2012practical, alflen2020model}, our work is among the first to explore the role of taxonomy usage in this context. The finding that delayed taxonomy usage leads to the most diverse set of needs is particularly noteworthy and could influence how future requirement elicitation frameworks are designed. Additionally, our results challenge the assumption that structured taxonomies always improve elicitation—while they enhance categorization, their premature use may restrict the diversity of responses. 

Overall, our findings suggest that there is no universally superior elicitation method, but rather that the choice depends on the specific goals of the elicitation process. If a company aims to minimize resource expenditure, interviews should be conducted, as they are the most efficient method. If the goal is to identify as many explanation needs as possible, surveys are the best option. However, if ample resources are available, a combination of surveys and interviews is recommended to maximize both breadth and depth in requirement elicitation. Based on our results, companies should adopt a hybrid approach, leveraging interviews for efficiency, surveys for scalability, and taxonomies strategically to maximize the completeness and quality of collected requirements.

\subsection{Threats to Validity}
\label{sec:validiteat}

The following section applies the "Threats to Validity" as described by Wohlin~\cite{wohlin2012experimentation} to the content of this work. These threats are categorized into \textit{construct}, \textit{internal}, \textit{conclusion}, and \textit{external} validity.

\subsubsection{Construct Validity}
One potential threat is the reliance on distinct needs as the primary metric for effectiveness. While this prevents inflation from redundant responses, it may overlook nuances in how explanation needs are articulated. Additionally, the introduction of a taxonomy could have influenced participants' responses by steering them toward predefined categories rather than capturing more spontaneous needs.

The categorization of needs was performed by a single requirements engineer who consulted another expert in cases of uncertainty. However, inter-rater reliability was not explicitly measured, which could impact reproducibility. While a second requirements engineer was consulted in cases of uncertainty, the lack of a systematic inter-rater agreement assessment means coder subjectivity may still have influenced the final coding of needs. The classification of unique needs also carries the risk of misclassification, potentially affecting the accuracy of the results.

A potential threat is that \enquote{no taxonomy usage} and \enquote{delayed taxonomy usage} were not examined in separate studies, making the \enquote{no taxonomy usage} data a subset of the \enquote{delayed taxonomy usage} data. While this introduces dependencies, it also ensures that comparisons are not influenced by external factors like participant differences or environmental variations, providing a controlled insight into the impact of taxonomy introduction.

\subsubsection{Internal Validity}
A key limitation is the difference in data collection conditions: focus groups were conducted in person, while interviews and surveys took place online. This discrepancy may have influenced engagement, discussion flow, and facilitator interaction. Additionally, the two focus groups differed in composition—one consisted of participants who knew each other well, while the other was composed of individuals with less prior interaction, potentially affecting group dynamics.

Another threat concerns facilitation bias: The specific role and behavior of facilitators in interviews and focus groups were not systematically documented or controlled, which may have influenced the responses and depth of elicited needs.

The survey exhibited a high dropout rate, particularly after the software usage questions and the introduction to explainability. This suggests that some participants found the survey too complex or time-consuming, leading to potential self-selection bias. Responses may therefore predominantly reflect the views of more motivated participants. In addition, since the survey was conducted remotely, we cannot determine whether participants remained actively engaged throughout the entire session or if they left the survey open while attending to other tasks. However, this reflects a realistic scenario for online surveys in practice, where participants complete them at their own pace and with varying levels of focus, making this threat minimal in real-world applications.

Variations in session duration could have influenced participant attention, particularly in longer sessions where cognitive fatigue may have played a role. However, the durations of all three elicitation methods were comparable.

\subsubsection{Conclusion Validity}
The participant distribution across methods was imbalanced, with 188 survey respondents compared to 18 interview participants and 12 in focus groups, which affects the direct comparability of methods due to differences in statistical power. However, to ensure a fair evaluation of efficiency and effectiveness, the data was normalized, allowing for a more balanced comparison across elicitation methods.

No statistical tests were conducted to assess the significance of differences observed between elicitation methods. As a result, reported differences should be interpreted as descriptive rather than statistically confirmed.

While taxonomy usage was controlled in a structured manner, participants’ prior knowledge of explainability concepts was not explicitly measured. Differences in familiarity could have influenced how well participants articulated their needs. Additionally, the manual categorization of needs introduces potential subjectivity, as inter-rater reliability was not systematically assessed.

\subsubsection{External Validity}
The study was conducted within a single German company using personnel management software. The findings may not fully generalize to other domains, such as security-critical or AI-driven applications, where explanation needs could differ. Moreover, as the study focused on human resources software, the specific domain context may have biased the types of explanation needs collected. The influence of domain characteristics was acknowledged but not analyzed in depth. Furthermore, all participants were employees of the same company, limiting the diversity of perspectives.

Only three elicitation methods—focus groups, interviews, and surveys—were examined. Alternative approaches, such as ethnographic studies, participatory design, or observational techniques, were not included. Additionally, the high survey dropout rate suggests that online elicitation methods require careful design to maintain engagement, particularly for complex topics like explainability.

The role of taxonomies in elicitation was explored, but their impact beyond this stage remains unclear. Further research is needed to determine whether taxonomies support prioritization, validation, and implementation of explainability requirements in software development.

\subsection{Future Work}
\label{sec:future-work}

Our study provides valuable insights into the efficiency and effectiveness of different elicitation methods for explainability requirements. However, several avenues for future research remain open.

To confirm the generalizability of our results, similar studies should be conducted in different software domains (e.g., safety-critical systems, AI-driven applications) and across companies in different countries. This would help determine whether cultural or industry-specific factors influence the optimal choice of elicitation methods. Additionally, further research could explore automated or semi-automated approaches to support the elicitation process, such as AI-driven questionnaires or natural language processing techniques to refine and categorize explanation needs dynamically.

While our study focused on focus groups, interviews, and surveys, future research could investigate additional elicitation methods such as workshops, observations, user diaries, ethnographic studies, or participatory design approaches. These methods might offer alternative advantages in different organizational settings, especially when explanation needs are complex and context-dependent. Furthermore, an extended study could examine how group size, facilitator involvement, or interaction styles influence the effectiveness of these elicitation methods.

Another promising direction is to broaden the scope beyond explainability requirements and apply a similar methodology to other NFRs, such as usability, security, or transparency. Investigating whether elicitation methods yield different patterns for various NFRs could help refine best practices for software requirements engineering in general.

Additionally, the role of taxonomy usage in elicitation could be further explored. Our results suggest that delayed taxonomy usage is particularly effective, making it worthwhile to refine the taxonomy further by adding new categories or clarifying existing ones. Future studies could also investigate whether taxonomies not only aid in collecting requirements but also support requirements engineers and developers in structuring, prioritizing, and implementing explanations more effectively. Furthermore, using different taxonomies tailored to specific domains or software types could provide insights into their adaptability across various projects.

Our dataset includes demographic factors such as job role, age, and software usage frequency, which were not analyzed in this study. Future research could explore correlations between user characteristics and explanation needs, identifying patterns that could refine the selection of elicitation methods for different stakeholder groups.

Finally, our findings suggest that taxonomy usage impacts the diversity of collected needs, but the long-term impact of taxonomy usage in software development remains unclear. Future work could explore how taxonomies influence requirement validation, prioritization, and integration into agile development workflows. Investigating how teams adopt taxonomies beyond the elicitation phase—such as in design, implementation, and evaluation—could provide further guidance on their overall value in software engineering.

\section{Conclusion}
\label{sec:conclusion}
This study investigated the efficiency and effectiveness of different elicitation methods for collecting explainability requirements. We conducted a case study at a large German IT consulting firm, comparing focus groups, interviews, and surveys. In this context, we examined the impact of using a taxonomy for categorizing explanation needs, evaluating whether introducing it (1) at the beginning, (2) after an initial elicitation phase, or (3) not at all influenced the number and distribution of collected requirements.

Our results showed that interviews were the most efficient method, yielding the highest number of distinct needs per participant per time spent. Surveys proved to be the most effective in capturing a large volume of explanation needs, but also had the highest redundancy, with many needs being stated multiple times. Focus groups, while generating meaningful discussions, were the least efficient in terms of needs collected per effort spent. 
These findings indicate that there is no universally superior elicitation method; instead, the choice depends on the objectives and resource constraints of an organization. If efficiency is the priority, interviews should be used. If the goal is to collect the most comprehensive set of explanation needs, surveys are the best option. When ample resources are available, a hybrid approach combining interviews and surveys is recommended to maximize both depth and breadth in requirements elicitation. 

Notably, different elicitation methods captured different types of explanation needs, with little overlap between them. In the delayed taxonomy condition, 105 needs were exclusive to interviews, 336 to surveys, and only 25 were shared between both methods. The overlap was even smaller for focus groups, with just 1–2 needs shared across methods. This suggests that a combination of methods is necessary for comprehensive coverage.

Future work should explore how automated approaches could further support the elicitation process, particularly in integrating taxonomies more seamlessly into different elicitation methods.

\section*{Acknowledgment}
This work was funded by the Deutsche Forschungsgemeinschaft (DFG, German Research Foundation) under Grant No.: 470146331, project softXplain (2022-2025).

\bibliographystyle{IEEEtran}
\bibliography{references.bib}

\end{document}